\documentclass[conference,a4paper]{IEEEtran}

\usepackage[utf8]{inputenc}
\usepackage[T1]{fontenc}
\usepackage{url}              % provides \url{...}
\usepackage{cite}             % improves presentation of citations

\usepackage[cmex10]{amsmath}  % Use the [cmex10] option to ensure complicance
\usepackage{mleftright}       % fix to wrong spacing of \left-,
\mleftright                   % \middle- \right-commands

\usepackage{graphicx}         % provides \includegraphics{...} to
\usepackage{booktabs}         % fixes poor spacing in tables and
\newcommand {\mymarginpar}[1]{\marginpar{#1}}
\renewcommand {\marginpar}[1]{}

\def\_{\rule{.3em}{.15ex}}      % Get underscore by typing \_.

\newcommand{\ls}[1]
   {\dimen0=\fontdimen6\the\font
    \lineskip=#1\dimen0
    \advance\lineskip.5\fontdimen5\the\font
    \advance\lineskip-\dimen0
    \lineskiplimit=.9\lineskip
    \baselineskip=\lineskip
    \advance\baselineskip\dimen0
    \normallineskip\lineskip
    \normallineskiplimit\lineskiplimit
    \normalbaselineskip\baselineskip
    \ignorespaces
   }
\newcommand {\bearn}{\begin{eqnarray*}}
\newcommand {\eearn}{\end{eqnarray*}}
\newcommand {\barr}{\begin{array}}
\newcommand {\earr}{\end{array}}

\newcommand {\N}{{\cal N}}

\newtheorem{definition}{Definition}
\newtheorem{property}[definition]{Property}
\newtheorem{proposition}[definition]{Proposition}
\newtheorem{lemma}[definition]{Lemma}
\newtheorem{theorem}[definition]{Theorem}
\newtheorem{corollary}[definition]{Corollary}
\newtheorem{example}{Example}
\newtheorem{remark}[definition]{Remark}

\newcommand {\benum} {\begin{enumerate}}
\newcommand {\eenum} {\end{enumerate}}

\newcommand {\bdesc} {\begin{description}}
\newcommand {\edesc} {\end{description}}

\newcommand {\bfig}[2] {\begin{figure}
  \centering
  \includegraphics[width=#2]{#1}}
\newcommand {\brotatefig}[2] {\begin{figure}[htbp]
                        \centerline {
                         \epsfig{figure={#1},clip=,angle=-90,width={#2}}}}
\newcommand {\bfigfirst}[2] {\begin{figure}[h]
                        \centerline {
                        \setlength{\epsfxsize}{#2}
                        \epsffile{#1}}}
\newcommand {\efig}[2]{ \caption{#2}
                        \label{fig:#1}
                        \end{figure}
                        \mymarginpar{fig:#1}}
\newcommand {\erotatefig}[2]{ \caption{#2}
                        \label{fig:#1}
                        \end{figure}
                        \mymarginpar{fig:#1}}
\newcommand {\rfig}[1]{Figure \ref{fig:#1}}

\newcommand {\btab}[1]{
                       \begin{table}
                       \centering
                       \begin{tabular}{#1}}
\newcommand {\etab}[3] {
                       \end{tabular}
                       \caption[#3]{#2}
                       \label{tab:#1}
                       \end{table}
                       \mymarginpar{tab:#1}
                       \vspace{.1in}}

\newcommand {\btabular}[1]{\begin{center}
                       \begin{tabular}{#1}}
\newcommand {\etabular}{\end{tabular}
                       \end{center}}

\newcommand {\bdefin}[1]{\begin{definition}
                      \mymarginpar{def:#1}
                      \label{def:#1} }
\newcommand {\edefin}       {\end{definition}}

\newcommand {\bpro}[1]{\begin{property}
                      \mymarginpar{pro:#1}
                      \label{pro:#1} }
\newcommand {\epro}   {\end{property}}

\newcommand {\bprop}[1]{\begin{proposition}
                      \mymarginpar{prop:#1}
                      \label{prop:#1} }
\newcommand {\eprop}       {\end{proposition}}

\newcommand {\blem}[1]{\begin{lemma}
                      \mymarginpar{lem:#1}
                      \label{lem:#1} }
\newcommand {\elem}   {\end{lemma}}
\newcommand {\rlem}[1]{Lemma \ref{lem:#1}}

\newcommand {\bthe}[1]{\begin{theorem}
                      \mymarginpar{the:#1}
                      \label{the:#1} }
\newcommand {\ethe}   {\end{theorem}}
\newcommand {\rthe}[1]{Theorem \ref{the:#1}}

\newcommand {\bproof}{\noindent {\bf Proof.} \ }
\newcommand {\eproof} {\hfill \squares \\ \vspace{.3cm}}
\newcommand {\bcor}[1]{\begin{corollary}
                      \mymarginpar{cor:#1}
                      \label{cor:#1} }
\newcommand {\ecor}   {\end{corollary}}

\newcommand {\bax}[1]{\begin{axiom}
                      \mymarginpar{ax:#1}
                      \label{ax:#1} }
\newcommand {\eax}       {\vspace{-.1in} \end{axiom}}

\newcommand {\bex}[2]{\vspace{.1in}
                      \begin{example}
                      \mymarginpar{ex:#1}
                       {\bf #2}
                      \label{ex:#1} }
\newcommand {\eex}       {\end{example} \vspace{.3cm} }
\newcommand {\rex}[1]{Example \ref{ex:#1}}

\newcommand {\brem}[1]{\begin{remark}
                      \mymarginpar{rem:#1}
                      \label{rem:#1} \em }
\newcommand {\erem}   {\end{remark}}

\newcommand {\beq}[1]{\mymarginpar{eq:#1}
                      \begin{equation}
                      \label{eq:#1} }

\newcommand {\beqno}[1]{\mymarginpar{eq:#1}
                      \begin{eqnarray}
                      \nonumber}

\newcommand {\eeq}       {\end{equation}}
\newcommand {\eeqno}       { && \end{eqnarray}}
\newcommand {\req}[1]{(\ref{eq:#1})}

\newcommand {\bear}[1]{\mymarginpar{eq:#1}
                       \begin{eqnarray}
                       \label{eq:#1} }

\newcommand {\bearno}[1]{\mymarginpar{eq:#1}
                       \begin{eqnarray}
                       \nonumber}

\newcommand {\eear}{\end{eqnarray}}
\newcommand {\eearno}{\end{eqnarray}}
\newcommand {\bsel}{\left \{ \begin{array}{cl}}
\newcommand {\esel}{\end{array} \right.}

\newcommand {\bmat}[1]{\left [ \begin{array}{#1}}
\newcommand {\emat}{\end{array} \right ]}
\newcommand {\bsec}[2]{\mymarginpar{sec:#2}
                       \section{#1}
                       \label{sec:#2} }

\newcommand {\rsec}[1]{Section \ref{sec:#1}}

\newcommand {\bsubsec}[2]{\mymarginpar{sec:#2}
                       \subsection{#1}
                       \label{sec:#2} }

\def\R{I\kern-0.30em R}
\def\N{I\kern-0.30em N}
\def\P{I\kern-0.30em P}
\newcommand\squares{\vrule height6pt width7pt depth1pt}

\def\ex{{\bf\sf E}}
\def\pr{{\bf\sf P}}

\begin{document}

\title{Degree-degree Correlated Low-density Parity-check Codes Over a Binary Erasure Channel}

\iffalse
\author{%
  \IEEEauthorblockN{Anonymous Authors}
  \IEEEauthorblockA{%
    Please do NOT provide authors' names and affiliations\\
    in the paper submitted for review, but keep this placeholder.\\
    ISIT23 follows a \textbf{double-blind reviewing policy}.}
}
\fi

\author{Hsiao-Wen Yu, Cheng-En Lee, Ruhui Zhang, 
	Cheng-Shang Chang,~\IEEEmembership{Fellow,~IEEE}and Duan-Shin Lee,~\IEEEmembership{Senior Member,~IEEE}\\
	Institute of Communications Engineering\\
	National Tsing Hua University \\
	Hsinchu 30013, Taiwan, R.O.C. \\
	Email: yuhw9817@gapp.nthu.edu.tw; benny\_110065508@gapp.nthu.edu.tw; huibrana@gapp.nthu.edu.tw;\\
	cschang@ee.nthu.edu.tw; lds@cs.nthu.edu.tw\\
}

\maketitle

\begin{abstract}
Most existing works on analyzing the performance of a random ensemble of low-density parity-check (LDPC) codes assume that the degree distributions of the two ends of a randomly selected edge are independent.
In the paper, we take one step further and consider ensembles of LDPC codes with degree-degree correlations.
For this, we propose two methods to construct an ensemble of degree-degree correlated LDPC codes. We then
derive a system of density evolution equations for such degree-degree correlated LDPC codes over a binary erasure channel (BEC). By conducting extensive numerical experiments, we show how the degree-degree correlation affects the performance of LDPC codes. Our numerical results show that LDPC codes with negative degree-degree correlation could improve the maximum tolerable erasure probability.
% {\color{red}for the first method and construct LDPC codes with an arbitrary bivariate degree-degree distribution to gain the performance improvement of LDPC codes for the second method.}
 Moreover, increasing the negative degree-degree correlation could lead to better unequal error protection (UEP) design.
\end{abstract}

%\begin{IEEEkeywords}
%{\bf \textit{Index Terms---}Low-density parity-check codes, unequal error protection.}
%\end{IEEEkeywords}
%\keywords{random access, stability, successive interference cancellation}

\section{Introduction}
\label{sec:introduction}

Low-density parity-check (LDPC) codes, first introduced by R. Gallager in 1962 \cite{gallager1962low}, have been widely used in practice, including the 5G new radio wireless communication standard (see, e.g., \cite{bae2019overview,thi2021low}) and the flash-memory systems (see, e.g., \cite{lee2017ldpc,fang2021irregular}). LDPC codes are linear block codes that can be characterized by either a generator matrix or  a parity-check matrix. It is well-known that there is a one-to-one mapping between the generator matrix and the parity-check matrix (see, e.g., Section II.B of the survey paper \cite{bonello2010low}).
In particular, for an $(n,k)$-LDPC code, there are $k$ information bits in a codeword of $n$ bits. The $n \times k$ generator matrix then maps the $k$ information bits to the $n$ codeword bits by multiplying the vector of the information bits with the generator matrix. On the other hand, the $n \times (n-k)$ parity-check matrix can be used for checking whether an $n$-bit binary vector is a legitimate codeword and decoding the $k$ information bits from that codeword.

The $n \times (n-k)$ parity-check matrix associated with an $(n,k)$-LDPC code can be represented by a bipartite graph, commonly known as the Tanner graph \cite{tanner1981recursive}. For such a bipartite graph, there are $n$ variable nodes on one side and $n-k$ check nodes on the other side. A check node represents a constraint on the values of variable nodes connected to that check node. 
In particular, for binary-valued variables, the value of a check node is simply the checksum of the values of the variable nodes connected to that check node. For a legitimate codeword, the checksums of the $n-k$ check nodes are all 0's. In other words, multiplying a codeword with the parity-check matrix yields a vector with all 0's.

Performance analysis of LDPC codes over a binary erasure channel (BEC) has been studied extensively in the literature (see e.g., \cite{luby1998analysis,luby1998analysisb,shokrollahi1999new,richardson2001capacity}). For a BEC with the erasure probability $\delta$, each bit in a codeword is erased {\em independently} with probability $\delta$. To recover the erased code bits, one can apply an iterative decoding (peeling) algorithm described below.
%One first adds the values of the non-erased variable nodes
%(in $GF(2)$)
%to the check nodes connected to them. Then remove all these non-erased nodes and their edges. The remaining bipartite graph only consists of erased variable nodes.
%The process for updating the values of nodes in a bipartite graph after removing some nodes is as follows: 
First, the values of the non-erased variable nodes are added to the check nodes connected to them. Subsequently, these non-erased nodes and their edges are removed. The remaining bipartite graph consists only of erased variable nodes.
%An erased variable node connected to a check node with degree 1 can be decoded by the value of that check node. Once decoded, it becomes  a non-erased variable node and can be removed in the same way as the initial non-erased variable nodes. The process is repeated until no more erased variable nodes can be decoded.
An erased variable node connected to a check node with degree 1 can be decoded by the value of that check node. Upon successful decoding, the erased variable node becomes a non-erased node and can be removed similarly to the original non-erased nodes. Repeat this process until no more erased variable nodes can be decoded.

A probabilistic framework for analyzing the iterative  algorithm is known as the {\em density evolution} method \cite{luby1998analysisb,shokrollahi1999new,richardson2001capacity}.
Such a method tracks the probability that the variable end of a randomly selected edge is not decoded after each iteration. The evolution of such a probability is characterized by the degree distribution of a randomly selected variable node and that of a randomly selected check node.
The density evolution method can also be extended to provide unequal error protection of code bits \cite{rahnavard2006new,rahnavard2007unequal,rahnavard2007rateless,chen2020analysis,zhao2022duplicated}.
For the setting with unequal error protection, code bits are classified into several classes, and the
density evolution method tracks the probability that the variable end of a randomly selected edge {\em in each class} is not decoded after each iteration.

One common assumption used in the density evolution method for LDPC codes is that the degree distributions of the two ends of a randomly selected edge are {\em independent}. \iffalse As such, the density evolution equations can be easily characterized by the two degree distributions of the variable nodes and the check nodes.\fi It seems that the effect of the degree-degree correlation of a randomly selected edge on LDPC codes has not been studied. One possible reason is that LDPC codes are  (nearly) capacity-achieving codes when the degree distributions are optimized \cite{mackay1997near,mackay1999good,richardson2001design}. However, the optimized degree distributions are in general highly irregular \cite{luby1997practical,shokrollahi1999new,richardson2001design,giddens2021enumeration}, and the throughput performance of regular LDPC codes is worse than irregular LDPC codes. The relaxation from {\em independent} degree distributions to a general bivariate distribution allows us to improve the performance of LDPC codes in the literature.

The main objective of this paper is to study the effect of the degree-degree correlation of a randomly selected edge on LDPC codes over a BEC. We summarize our contributions as follows:
\begin{itemize}
%[leftmargin=*]
\item {\bf Construction}: As a generalization of the configuration model (with independent degree distributions) \cite{Newman2010}, we propose two constructions of random LDPC codes (bipartite graphs) with degree-degree correlations.
The first construction, called the block construction in this paper, is an extension of the construction for uni-partite graphs in \cite{lee2019generalized}.
For such a model, we derive a closed-form expression of the degree-degree bivariate distribution of the two ends of a randomly selected edge. The second construction, called the general construction, is more general than the first one,
and it can construct random LDPC codes with a specified degree-degree bivariate distribution.
By classifying edges with the same degrees in the variable end and the check end into one specific type of edges,
the second construction can be viewed as a special class of the multi-edge type LDPC codes in \cite{richardson2002multi} that automatically satisfy the stub count constraints.

\item {\bf Analysis}: It is known that the density evolution method is generally difficult to be applied to the multi-edge type LDPC codes as the densities of various types are convolved. However, by using the degree-degree bivariate distribution, we can average over the (convolved) densities of various {\em edge} types into a single {\em node} type. As such, we derive a system of density evolution equations over a BEC\iffalse binary erasure channel\fi. Such an extension covers the independent case as a special case.  From these systems of density evolution equations, we derive lower bounds on the maximum tolerable erasure probability (such that every code bit is recovered with high probability).

\item {\bf Effect of correlation}: By conducting extensive numerical experiments, we show the effect of the degree-degree correlation on the performance of LDPC codes. Our numerical results show that optimizing {\em negative} correlation can achieve a much higher  maximum tolerable erasure probability than LDPC codes with independent degree distributions. This implies that {\em partially regular} LDPC codes \cite{rahnavard2007unequal}  with degree-degree correlation might be good enough for practical use.

\item {\bf Performance improvement}: By taking the degree-degree correlation into consideration, we can further improve the best-known result for the maximum tolerable erasure probability in the literature. In particular, we show that under the same marginal degree distributions of the variable ends and the check ends in \cite{shokrollahi2000design}, the maximum tolerable erasure probability can be extended from 0.49553 to 0.49568.

 \end{itemize}

\bsec{Construction of degree-degree correlated random LDPC codes}{construction}
Instead of assuming that the degree of the variable end and the degree of the check end of a randomly selected edge are independent in \cite{shokrollahi1999new}. \iffalse \req{const2222},\fi In this section, we construct a random bipartite graph with a specific degree-degree correlation.
%In this section, we first briefly review the density evolution method for LDPC codes with indepedent degree distribution. The random $(n,k)$-LDPC code in \cite{shokrollahi1999new} is constructed by specifying the degree distributions of the variable nodes and check nodes.
Suppose that there are $n$ variable nodes and $n-k$ check nodes. Let $p_X(x)$ (resp. $p_Y(y)$) be the degree distribution of a randomly selected variable node $X$ (resp. check node $Y$).
%, and $p_Y(y)$ be the degree distribution of a randomly selected check node $Y$.
Also, let
$\ex[X]= \sum_x  xp_X(x)$ (resp. $\ex[Y]=\sum_y yp_Y(y)$) be the average degree of a randomly selected variable node (resp. check node).
%, and $\ex[Y]=\sum_y yp_Y(y)$ be the average degree of a randomly selected check node.
%$G=n/(n-k)$ be the normalized load.
Following the argument for the configuration model in \cite{Newman2010}, we generate $n p_X(x)$ variable nodes with degree (stub) $x$ and $(n-k) p_Y(y)$ check nodes with degree (stub) $y$. Then the total number of stubs for variable nodes is $\sum_x nx p_X(x)$, and the total number of stubs for check nodes is $\sum_y (n-k)yp_Y(y)$.
Suppose that
\beq{const1111}
n\ex[X] =(n-k)\ex[Y].
\eeq
%Then we can randomly select one stub from the variable nodes and another stub from the check nodes and connect these two stubs to form an edge. Repeating the process yields a bipartite graph with $n$ variable nodes and $n-k$ check nodes.
In this paper, we let
\beq{const1155}
G=\frac{n}{n-k}=\frac{\ex[Y]}{\ex[X]}
\eeq
when the identity in \req{const1111} is satisfied.

\iffalse
Let $X_e$ (resp. $Y_e$) be the degree of the variable end (resp. check end) of a randomly selected edge from the bipartite graph.
Then
\bear{const2222}
\pr (X_e=x, Y_e=y)&=& \pr (X_e=x) \pr(Y_e=y) \nonumber\\
&=&\frac{xp_X(x)}{\ex[X]}\frac{yp_Y(y)}{\ex[Y]}.
\eear
\fi

\bsubsec{Block construction}{block}

%Instead of assuming that the degree of the variable end and the degree of the check end of a randomly selected edge are independent in \req{const2222}, in this section we construct a random bipartite graph with a specific degree-degree correlation.
In this section, our construction for such a bipartite graph is similar to the block construction of degree-degree correlated random networks in \cite{lee2019generalized}. %We assume  that there are $n$ variable nodes and $n-k$ check nodes. Let $p_X(x)$ be the degree distribution of a randomly selected variable node $X$, and $p_Y(y)$ be the degree distribution of a randomly selected check node $Y$.
%Also, let
%$\ex[X]= \sum_x  xp_X(x)$ be the average degree of a randomly selected variable node and
%$\ex[Y]=\sum_y yp_Y(y)$ be the average degree of a randomly selected check node.
Assume that \req{const1111} holds. We arrange the $n \ex[X]$ stubs of the variable nodes in descending order and
partition the $n \ex[X]$ stubs into $b$ blocks evenly, each with $n\ex[X]/b$ stubs. In each block of $n\ex[X]/b$ stubs, we {\em randomly} select $q n\ex[X]/b$ stubs as type 1 stubs, where the parameter $q$ is a design parameter  for controlling the degree-degree correlation. The remaining $(1-q) n\ex[X]/b$ stubs in that block are classified as type 2 stubs.
Similarly, we arrange the $(n-k) \ex[Y]$ stubs of the check nodes in descending order and
partition the $(n-k) \ex[Y]$ stubs into $b$ blocks evenly, each with $(n-k) \ex[Y]/b$ stubs. In each block of $(n-k) \ex[Y]/b$ stubs, we {\em randomly} select $q (n-k) \ex[Y]/b$ stubs as type 1 stubs
and the remaining $(1-q) (n-k) \ex[Y]/b$ stubs as type 2 stubs.
These two types of stubs will be connected differently to form degree-degree correlation.

Consider a permutation $\pi$ of $\{1,2, \ldots, b\}$. For $i=1,2, \ldots, b$, we randomly select a type 1 stub from the $i^{th}$ block of variable nodes and another type 1 stub from the $\pi(i)^{th}$ block of  check nodes and connect these two stubs to form an edge. Repeat the process until all the type 1 stubs are connected. For type 2 stubs, the connection is independent,
%the same as that in \rsec{randomLDPC}
i.e., we randomly select a type 2 stub of variable nodes and another type 2 stub  of  check nodes and connect these two stubs to form an edge.
Let $X_e$ (resp. $Y_e$) be the degree of the variable end (resp. check end) of a randomly selected edge from the bipartite graph.
It is clear that the marginal distribution of the degree of the variable end (of a randomly selected edge) is
\beq{const2222c}
\pr (X_e=x)=\frac{xp_X(x)}{\ex[X]},
\eeq
and the marginal distribution of the degree of the check end (of a randomly selected edge) is
\beq{const2222d}
\pr(Y_e=y) =\frac{yp_Y(y)}{\ex[Y]}.
\eeq

Assume that all the nodes with the same degree are in one common block. Let $S_X(i)$ (resp. $S_Y(i)$), $i=1,2, \ldots, b$, be the set of degrees in the $i^{th}$ block of variable nodes (resp. check nodes).
Now we compute the conditional probability $\pr (Y_e=y | X_e=x)$.
Suppose that $x \in S_X(i)$ for some $i$. If $y \not \in S_Y(\pi(i))$, then an edge with degree $x$ at the variable end and degree $y$ at the check end can only be formed by type 2 stubs. The number of type 2 stubs of the check nodes with degree $y$ is $(1-q)(n-k) yp_Y(y)$, and the total number of type 2 stubs is $(1-q)(n-k)\ex[Y]$. The probability that a stub of the variable nodes is of type 2 is $1-q$.
Thus, for $x \in S_X(i)$ and $y \not \in S_Y(\pi(i))$,
\bear{joint1111}
\pr (Y_e=y | X_e=x)&=&(1-q) \frac{(1-q)(n-k) yp_Y(y)}{(1-q)(n-k)\ex[Y]} \nonumber\\
&=&\frac{(1-q)yp_Y(y)}{\ex[Y]}.
\eear
If $y  \in S_Y(\pi(i))$, then an edge with degree $x$ at the variable end and degree $y$ at the check end can also be formed by type 1 stubs.
The number of type 1 stubs of the check nodes with degree $y$ is $q(n-k) yp_Y(y)$, and the total number of type 1 stubs in the $\pi(i)^{th}$ block of check nodes is $q(n-k)\ex[Y]/b$. The probability that a stub of the variable nodes is of type 1 is $q$.
Adding the probability formed by type 1 stubs in \req{joint1111}, we have for $x \in S_X(i)$ and $y \in S_Y(\pi(i))$,
\bear{joint2222}
&&\pr (Y_e=y | X_e=x)\nonumber\\
&=&q \frac{q(n-k) yp_Y(y)}{q(n-k)\ex[Y]/b} + \frac{(1-q)yp_Y(y)}{\ex[Y]} \nonumber\\
&=&\frac{(bq+(1-q))yp_Y(y)}{\ex[Y]}.
\eear
\iffalse
Since
\beq{joint3333}
\pr(X_e=x)=\frac{xp_X(x)}{\ex[X]},
\eeq
\fi
Then, we have from \req{const2222c}, \req{const2222d}, \req{joint1111}, and \req{joint2222} that for $x \in S_X(i)$,
\bear{joint}
&&\pr (X_e=x, Y_e=y)\nonumber\\
&&=\left\{\begin{array}{ll}
(bq+(1-q)) p_{X_e}(x) p_{Y_e}(y),
& y \in S_Y(\pi(i)),\\
(1-q) p_{X_e}(x) p_{Y_e}(y), &
 y \not \in S_Y(\pi(i)). \\
\end{array}\right. \nonumber\\
\eear

\bex{twodegrees}{(Two types of degrees)}
Suppose that there are $n/3$ variable nodes with degree $2d$,  and $2n/3$ variable nodes with degree $d$. Also,
there are $(n-k)/3$ check nodes with degree $2Gd$, and $2(n-k)/3$ check nodes with degree $Gd$,
where $G=n/(n-k)$ is a positive integer.
%In this example, we have $p_X(2d)=1/3$, $p_X(d)=2/3$, $p_Y(2Gd)=1/3$, and $p_Y(Gd)=2/3$.
Using the construction in \rsec{block}, we have $\ex[X]=4d/3$, $\ex[Y]=4Gd/3$ so that the condition in \req{const1111} is satisfied.
\iffalse Moreover, from \req{const2222c} and \req{const2222d}, it follows that
\bear{twod0000}
\pr (X_e=2d)&=&\frac{2d/3}{4d/3}=\frac{1}{2},\nonumber\\
\pr (X_e=d)&=&\frac{2d/3}{4d/3}=\frac{1}{2},\nonumber\\
\pr (Y_e=2Gd)&=&\frac{2Gd/3}{4Gd/3}=\frac{1}{2},\nonumber\\
\pr (Y_e=Gd)&=&\frac{2Gd/3}{4Gd/3}=\frac{1}{2}.
\eear \fi
Now we partition the stubs into two blocks, i.e., $b=2$. The first block consists of variable nodes with degree $2d$
(resp. check nodes with degree $2Gd$),
and the second block consists of variable nodes with degree $d$ (resp. check nodes with degree $Gd$).
For the permutation $\pi$ with $\pi(1)=2$ and $\pi(2)=1$,
we have from \req{const2222c} and \req{joint} that
\bear{twod1111}
\pr (Y_e=2Gd| X_e=2d)&=&\frac{1-q}{2}, \nonumber\\
\pr (Y_e=2Gd| X_e=d)&=&\frac{1+q}{2}, \nonumber\\
\pr (Y_e=Gd| X_e=2d)&=&\frac{1+q}{2}, \nonumber\\
\pr (Y_e=Gd| X_e=d)&=&\frac{1-q}{2} .
\eear
\eex

Note that the degree-degree correlation can be computed as follows:
\bear{cor1111}
\rho(X_e, Y_e) = \frac{\ex (X_eY_e)-\ex (X_e)\ex (Y_e)}{\sqrt{\mbox{Var}(X_e)}\sqrt{\mbox{Var}(Y_e)}} = -q.
\eear

\begin{figure}[ht]
	\centering
	\includegraphics[width=0.43\textwidth]{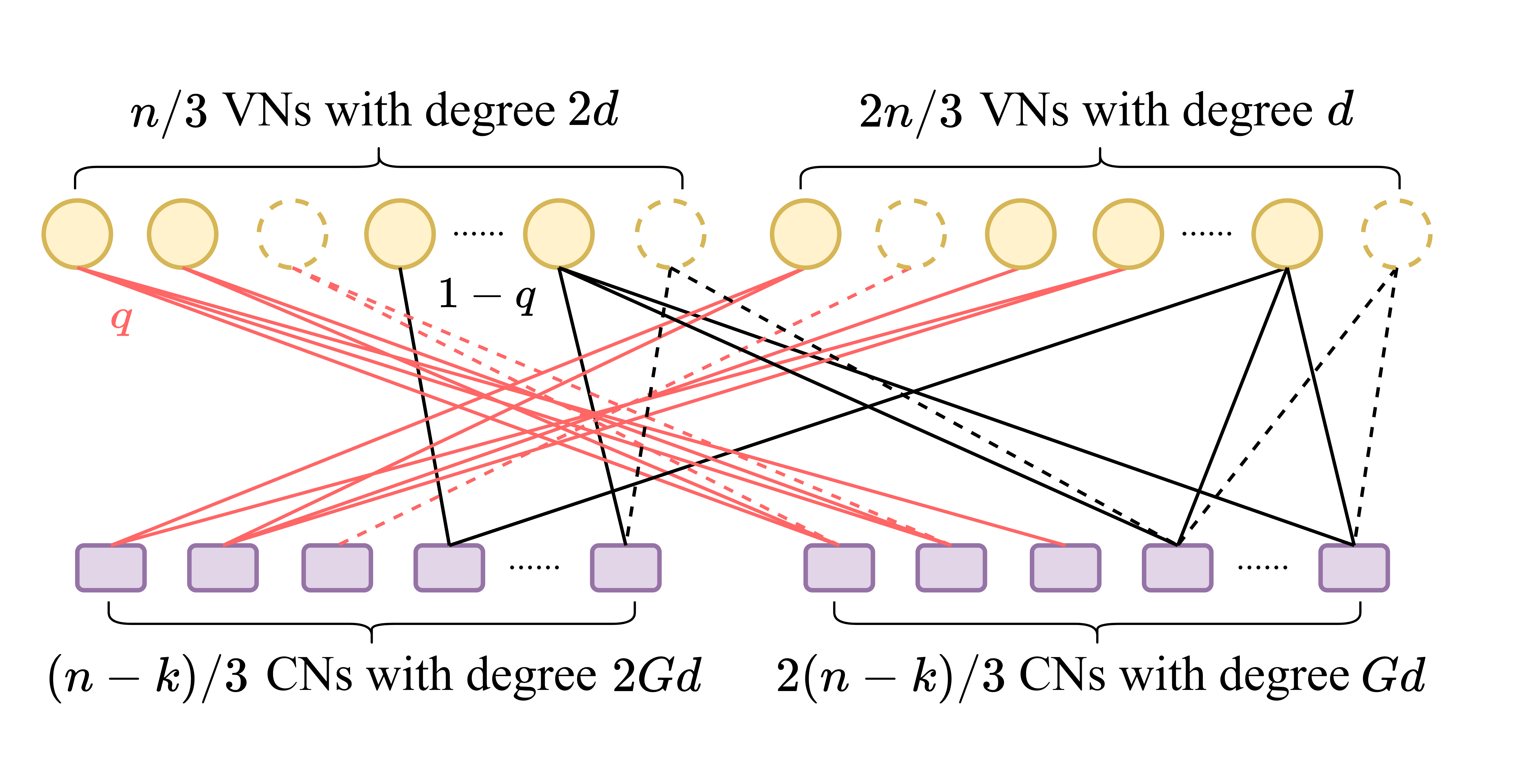}
	\caption{An illustration of the construction of the bipartite graph in \rex{twodegrees}, where the red (resp. black) edges are formed by type 1 (resp. 2) stubs.}
	\label{fig:qual}
\end{figure}

Thus, the degree-degree correlation is negative for the permutation $\pi$ with $\pi(1)=2$ and $\pi(2)=1$.
On the other hand, it is positively correlated for the permutation $\pi$ with $\pi(1)=1$ and $\pi(2)=2$.
%An illustration of the construction of different types of stubs is shown in \rfig{stub}, and
An illustration of the construction of the bipartite graph is shown in \rfig{qual}, where the red (resp. black) edges are formed by type 1 (resp. 2) stubs.

\iffalse
\begin{figure}[ht]
	\centering
	\includegraphics[width=0.43\textwidth]{stubs.png}
	\caption{An illustration of the construction of different types of stubs.}
	\label{fig:stub}
\end{figure}
\fi

\bsubsec{General construction}{general}

In this section, we propose a general construction from the degree-degree bivariate distribution $\pr (X_e=x, Y_e=y)$.
Given such a bivariate distribution, we first compute the two marginal distributions:
\beq{genc1111}
p_{X_e}(x)=\sum_y \pr (X_e=x, Y_e=y),
\eeq
and
\beq{genc2222}
p_{Y_e}(y)=\sum_x \pr (X_e=x, Y_e=y).
\eeq
Then we use the marginal distribution in \req{genc1111} and \req{const2222c} to compute the degree distribution of $X$:
\beq{genc3333}
p_X(x)=\frac{p_{X_e}(x)/x}{\sum_{x^\prime} p_{X_e}(x^\prime)/x^\prime}.
\eeq
Similarly, we have from the marginal distribution in \req{genc2222} and \req{const2222d} that
\beq{genc4444}
p_Y(y)=\frac{p_{Y_e}(y)/y}{\sum_{y^\prime} p_{Y_e}(y^\prime)/y^\prime}.
\eeq
From \req{genc3333} and \req{genc4444}, we have
\bear{genc5555}
\ex[X]&=&\frac{1}{\sum_{x^\prime} p_{X_e}(x^\prime)/x^\prime}, \\
\ex[Y]&=&\frac{1}{\sum_{y^\prime} p_{Y_e}(y^\prime)/y^\prime}.
\eear
Choose $n$ and $k$ so that the condition in \req{const1155} is satisfied.
It then follows from \req{const1155}, \req{const2222c}, and \req{const2222d} that
\bear{genc6666}
&&n x p_X(x) \pr (Y_e=y| X_e=x) \nonumber\\
&=&n \ex[X] p_{X_e}(x) \pr (Y_e=y| X_e=x) \nonumber\\
&=&n \ex[X] \pr (X_e=x, Y_e=y)\label{eq:genc6666c}\\
&=&(n-k) \ex[Y] p_{Y_e}(y) \pr (X_e=x| Y_e=y) \nonumber\\
&=&(n-k) yp_Y(y) \pr (X_e=x| Y_e=y).
\eear
Note that $n x p_X(x)$ is the number of stubs of variable nodes with degree $x$,
and $(n-k)y p_Y(y)$ is the number of stubs of check nodes with degree $y$. Among these stubs, we can randomly select
$n x p_X(x) \pr (Y_e=y| X_e=x)$ stubs from the variable nodes and $(n-k) yp_Y(y) \pr (X_e=x| Y_e=y)$ stubs from the check nodes, and connect them at random (as in the configuration model).
As the number of edges between a variable node with degree $x$ and a check node with degree $y$ is $n x p_X(x) \pr (Y_e=y| X_e=x)$ and the total number of edges is $n \ex[X]$,   the probability that
a randomly selected edge has degree $x$ in its variable end and degree $y$ in its check end
 is (cf. \req{genc6666c})
\beq{genc7777}
\pr (X_e=x, Y_e=y) .
\eeq
To provide further insight of this construction, we show an example with two types of degrees in Appendix \ref{sec:apptwodegreesb}.

It is of interest to point out the connections between our construction and the multi-edge type LDPC codes in \cite{richardson2002multi}. By classifying edges with degree $x$ in the variable end and degree $y$ in the check end into one specific type of edges, our construction is a special case of the multi-edge type LDPC codes in \cite{richardson2002multi}. However, as pointed out in \cite{richardson2002multi,liva2007protograph}, the density evolution method is generally difficult to be applied to the multi-edge type LDPC codes (as the densities of various types are convolved). By using the degree-degree bivariate distribution, we will show in \rsec{correlatedLDPC}
that one can average over the (convolved) densities of various {\em edge} types into a single {\em node} type (that only depends on its node degree). Thus, the computational complexity of the density evolution method can be greatly reduced.

\bsec{Density evolution in correlated LDPC codes}{correlatedLDPC}

In this section, we extend the density evolution analysis for LDPC codes with degree-degree correlations.
It is known that the density evolution analysis is generally difficult to be applied to the multi-edge type LDPC codes as the densities of various types are convolved. However, by using the conditional probabilities derived in the previous section, we can average over the (convolved) densities of various {\em edge} types into a single {\em node} type.

%Analogous to the density evolution analysis in correlated LDPC codes, we can extend those \iffalse \req{andor1111a} and \req{andor2222a} \fi by averaging over conditional probabilities.
%{\color{red} Due to space limitation, the derivation of the density evolution method for irregular LDPC codes with independent degree distribution is not shown here, and it can be found in the full report in ArXiv with the same title. }
\iffalse
Let
%\beq{andor0033}
$\Lambda_X(z)=\sum_x z^x p_X(x)$
%\eeq
be the generating function of the degree distribution of $X$,
and
%\beq{andor0044}
$\lambda(z)=\sum_x z^{x-1} \frac{xp_X(x)}{\ex[X]}$
%\eeq
be the generating function of the {\em excess} degree distribution of $X$. Note that
%\beq{andor0055}
$\lambda(z)=\Lambda^\prime_X(z)/\ex[X].$
%\eeq
Similarly, let
%\beq{andor0066}
$\Lambda_Y(z)=\sum_y z^y p_Y(y)$
%\eeq
be the generating function of the degree distribution of $Y$,
and
%\beq{andor0077}
$\rho(z)=\sum_y z^{y-1} \frac{y p_Y(y)}{\ex[Y]}$
%\eeq
be the generating function of the {\em excess} degree distribution $Y$.
Note that
%\beq{andor0088}
$\rho(z)=\Lambda^\prime_Y(z)/\ex[Y].$
%\eeq
\fi

%(AND-OR decoder)
Now we derive the density evolution equations in the setting where $G$ is fixed and $n \to \infty$. Consider using a degree-degree correlated LDPC code over a binary erasure channel with the erasure probability $\delta$. Let $\alpha_x^{(i)}$ be
the probability that the variable end of a randomly selected edge with degree $x$ is not decoded after the $i^{th}$ iteration, and $\beta_y^{(i)}$  be the probability that the check end of a randomly selected edge with degree $y$ is not decoded after the $i^{th}$ iteration.
Analogous to the AND-OR argument in \cite{luby1998analysis,luby1998analysisb}, we have
\beq{andor1111c}
\alpha_x^{(i)}=\delta (\sum_y \beta_y^{(i)} \pr(Y_e=y |X_e=x))^{x-1}  .
\eeq
Clearly, $\alpha_x^{(0)}=\delta$ for all $x$ (as every variable bit is erased independently with probability $\delta$ through the erase channel). Similarly,
\beq{andor2222c}
\beta_y^{(i)}=1-(1-\sum_x \alpha_x^{(i-1)} \pr(X_e=x|Y_e=y))^{y-1}.
\eeq
Combining these two equations, we have a system of nonlinear recursive equations:
\bear{andor3333c}
&&\alpha_x^{(i)}\nonumber\\
&&=\delta \Big  (\sum_y \big (1-(1-\sum_{x^\prime} \alpha_{x^\prime}^{(i-1)} \pr(X_e=x^\prime|Y_e=y))^{y-1}\big )\nonumber\\
 &&\quad\pr(Y_e=y |X_e=x) \Big )^{x-1} ,
\eear
with $\alpha_x^{(0)}=\delta$ for all $x$.

Let $\gamma_x^{(i)}$ be the probability that a randomly selected variable node with degree $x$ is successfully decoded after the $i^{th}$ iteration. Note that a randomly selected variable node is not decoded after the $i^{th}$ iteration
if and only if the variable node is erased and all the {\em check} ends of its $x$ edges are not decoded after the $i^{th}$ iteration.
Thus,
\beq{andor7777}
\gamma_x^{(i)}=1-\delta(\sum_y \beta_y^{(i)} \pr(Y_e=y |X_e=x))^x.
\eeq
Let $\gamma^{(i)}$ be the probability that a randomly selected variable node  is successfully decoded after the $i^{th}$ iteration. Then
\bear{andor8888}
\gamma^{(i)} &=&\sum_x \gamma_x^{(i)} p_X(x)\nonumber\\
&=&1-\delta \sum_x (\sum_y \beta_y^{(i)} \pr(Y_e=y |X_e=x))^x p_X(x).\nonumber\\
\eear

As in \cite{shokrollahi1999new}, we are interested in the maximum tolerable erasure probability $\delta^*$ such that
for all $\delta < \delta^*$,
\beq{stab0000a}\lim_{i \to \infty}\gamma^{(i)} \to 1.
\eeq
Since the capacity of the BEC with the erasure probability $\delta$ is known to be $1-\delta$ \cite{Elias55},
we know that $\frac{k}{n} \le 1-\delta^*$. In view of \req{const1155}, we have
\beq{stab0066}
\delta^* \le \frac{1}{G}.
\eeq

In view of \req{andor8888} and \req{andor2222c}, a sufficient condition for \req{stab0000a} is
\beq{stab0000}\lim_{i \to \infty}\alpha_x^{(i)} \to 0
\eeq
 for all $x$.
In the following theorem, we show a lower bound for $\delta^*$. The proof can be found in 
Appendix \ref{sec:appproof}.
%the full report in ArXiv with the same title.

\bthe{stability}
Let $d_{v,\min}$ be the minimum degree of a variable node and $d_{c,\max}$ be the maximum degree of a check node.
If $d_{v,\min} \ge 2$ and $\delta <1/(d_{c,\max}-1)$, then
$\lim_{i \to \infty}\alpha_x^{(i)} \to 0$ for all $x$.
\ethe

\bsec{Numerical results}{numerical}

\bsubsec{Block construction}{blocknum}

In this section, we evaluate the performance of random LDPC codes generated by the block construction in \rsec{block}.
Consider using the randomly generated LDPC code in \rex{twodegrees} for a BEC with the erasure probability $\delta$.
\iffalse
Using \req{twod1111} in \req{andor1111c} and \req{andor2222c}, we derive the following density evolution equations:
\bear{twod2222}
\alpha_{d}^{(i)}&=&\delta (\frac{1-q}{2}\beta_{Gd}^{(i)}+\frac{1+q}{2}\beta_{2Gd}^{(i)})^{d-1}, \nonumber\\
\alpha_{2d}^{(i)}&=&\delta (\frac{1+q}{2}\beta_{Gd}^{(i)}+\frac{1-q}{2}\beta_{2Gd}^{(i)})^{2d-1}, \nonumber\\
\beta_{Gd}^{(i)}&=&1-(1-\frac{1-q}{2}\alpha_{d}^{(i-1)}-\frac{1+q}{2}\alpha_{2d}^{(i-1)} )^{Gd-1} , \nonumber\\
\beta_{2Gd}^{(i)}&=&1-(1-\frac{1+q}{2}\alpha_{d}^{(i-1)}-\frac{1-q}{2}\alpha_{2d}^{(i-1)} )^{2Gd-1} .\nonumber\\
\eear
{\color{red}
Then, using \req{twod2222} in \req{andor7777}, we have
\bear{twod2233}
\gamma^{(i)}_{d}&=&1-\delta(\frac{1-q}{2}\beta_{Gd}^{(i)}+\frac{1+q}{2}\beta_{2Gd}^{(i)})^d, \nonumber\\
\gamma^{(i)}_{2d}&=&1-\delta(\frac{1+q}{2}\beta_{Gd}^{(i)}+\frac{1-q}{2}\beta_{2Gd}^{(i)})^{2d}.
\eear
}
Note that the probability that a randomly selected variable node is successfully decoded after the $i^{th}$ iteration is
\bear{twod2244}
\gamma^{(i)}&=&1-\delta(\frac{1-q}{2}\beta_{Gd}^{(i)}+\frac{1+q}{2}\beta_{2Gd}^{(i)})^d \frac{2}{3}\nonumber\\
&&\quad -\delta (\frac{1+q}{2}\beta_{Gd}^{(i)}+\frac{1-q}{2}\beta_{2Gd}^{(i)})^{2d} \frac{1}{3}.
\eear
\fi
In this numerical experiment, we set  $G=3$, $d=3$.

\begin{figure}[ht]
	\centering
	\includegraphics[width=0.35\textwidth]{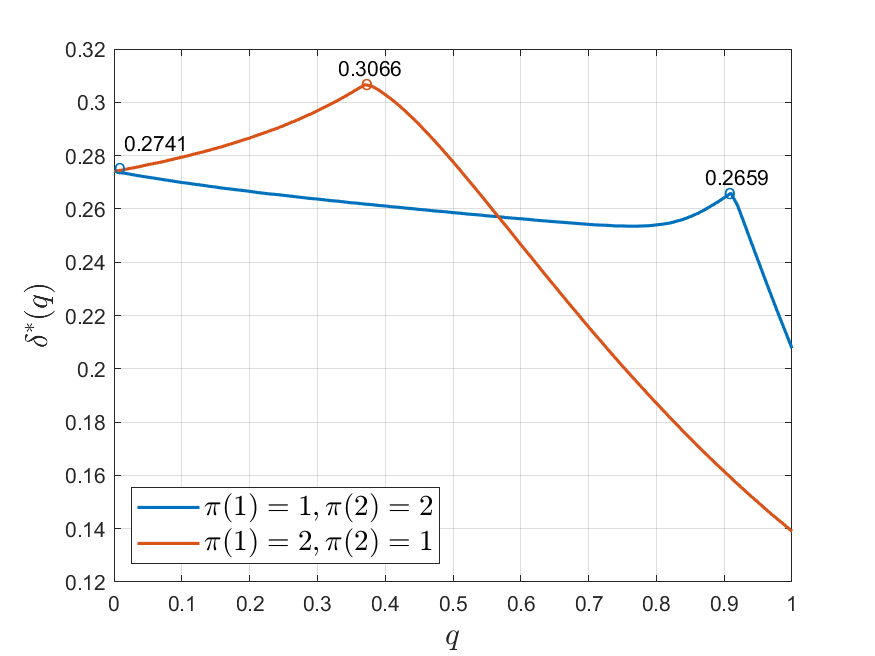}
	\caption{The maximum tolerable erasure probability $\delta^*(q)$ as a function of the parameter $q$
		(with the step size of 0.01) for two permutations: the negatively correlated one with $\pi(1)=2,\pi(2)=1$ (the orange curve), and the positively correlated one with $\pi(1)=1,\pi(2)=2$ (the blue curve).}
	\label{fig:mtep}
\end{figure}

In \rfig{mtep}, we
plot the maximum  tolerable erasure probability $\delta^*(q)$ as a function of the parameter $q$
(with the step size of 0.01) for two permutations: the negatively correlated one with $\pi(1)=2,\pi(2)=1$ (the orange curve), and the positively correlated one with $\pi(1)=1,\pi(2)=2$ (the blue curve). Recall that when $q=0$, it reduces to the independent case. As shown in \rfig{mtep}, adding a negative degree-degree correlation can lead to a much larger maximum tolerable erasure probability than that of the independent case. In particular, when $q=0.37$, we find
that the maximum tolerable erasure probability $\delta^*(q)$ is  0.3066, which is larger than $\delta^*(q) = 0.2741$ for $q = 0$. However, adding a positive degree-degree correlation does not improve the maximum  tolerable erasure probability in this numerical experiment.

To explain these numerical results, we note that  LDPC codes with positive degree-degree correlations tend to form a giant component in the bipartite graph (as nodes with large degrees tend to connect to nodes with large degrees), and that makes the decoding of the LDPC code difficult.
On the other hand,  LDPC codes with negative degree-degree correlations are less likely to form a giant component. As such, one can exploit the negative degree-degree correlation by increasing $q$ to improve the maximum  tolerable erasure probability. However, when $q$ is increased to 1, the bipartite graph gradually decouples into two separate bipartite graphs (from the block construction). Each has its own maximum  tolerable erasure probability. When $q$ is close to 1, the maximum  tolerable erasure probability is dominated by the bipartite graph with a smaller maximum  tolerable erasure probability.
Due to these two effects, the orange curve for the case with a negative degree-degree correlation is unimodal (with only one peak).

\iffalse
From \req{twod0000}, we obtain the expected value of $X_e$, $Y_e$, ${X_e}^2$, ${Y_e}^2$, and the standard deviation of $X_e$, $Y_e$
\bear{exp1111}
E(X_e)&=&\sum_x x\pr(X_e = x) = \frac{3d}{2} = 4.5,\nonumber\\
E(Y_e)&=&\sum_y y\pr(Y_e = y) = \frac{3Gd}{2} = 13.5,\nonumber\\
E({X_e}^2)&=&\sum_x x^2\pr(X_e = x) = \frac{5d^2}{2} = 22.5,\nonumber\\
E({Y_e}^2)&=&\sum_y y^2\pr(Y_e = y) = \frac{5G^2d^2}{2} = 202.5,\nonumber\\
\sigma_{X_e}&=&\sqrt{E({X_e}^2)-{E(X_e)}^2} = \frac{d}{2} = 1.5,\nonumber\\
\sigma_{Y_e}&=&\sqrt{E({Y_e}^2)-{E(Y_e)}^2} = \frac{Gd}{2} = 4.5.
\eear
Now we consider the expected value of the product $X_eY_e$ from \req{joint} that
\bear{exp2222}
E(X_eY_e)&=& \sum_x \sum_y xy\pr(X_e=x,Y_e=y)\nonumber\\
&=& d(Gd(1-q)\frac{1}{4}+2Gd(bq+(1-q))\frac{1}{4})\nonumber\\
&&+2d(Gd(bq+(1-q))\frac{1}{4}+2Gd(1-q)\frac{1}{4})\nonumber\\
&=&\frac{9Gd^{2}}{4}+\frac{Gd^{2}(4b-9)}{4}q = 60.75-6.75q.\nonumber\\
\eear
Then, the Pearson degree-degree correlation function of $X_e$ and $Y_e$ is
\bear{cor1111}
\rho(X_e, Y_e)&=&\frac{E(X_eY_e)-E(X_e)E(Y_e)}{\sigma_{X_e}\sigma_{Y_e}}\nonumber\\
&=&\frac{60.75-6.75q-4.5\times13.5}{1.5\times4.5}= -q.
\eear
We can see that $\rho(X_e, Y_e)$ is linear.
\fi

In Appendix \ref{sec:appsimulation}, we conduct extensive simulations to verify the effectiveness of the asymptotic results derived from the density evolution equations. We also show the degree's effect on the decoding probabilities of variable nodes. One significant finding from our experiments is that the degree-degree correlation  plays a critical  role in unequal error protection (UEP).
\iffalse
We also conduct extensive simulations to verify the effectiveness of the asymptotic results derived from the density evolution equations and show the degree-degree correlation plays a critical role in unequal error protection (UEP).
\fi
\bsubsec{General construction for performance improvement}{generalnum}

In this section, we show that the LDPC codes from the general construction in \rsec{general} can lead to further performance improvement.

In \cite{shokrollahi2000design}, Shokrollahi and Storn proposed
a construction of the LDPC code with the following (independent) bivariate distribution:
\beq{ss0000}
\pr (X_e=x, Y_e=y)=p_{X_e}(x)p_{Y_e}(y),
\eeq
where $p_{X_e}(2)=0.2633$, $p_{X_e}(3)=0.1802$, $p_{X_e}(7)=0.2700$, $p_{X_e}(30)=0.2865$, and $p_{Y_e}(8)=0.6341$, $p_{Y_e}(9)=0.3659$.
\iffalse
\bear{ss0011}
p_{X_e}(2)&=&0.26328, \nonumber\\
p_{X_e}(3)&=&0.18020, \nonumber\\
p_{X_e}(7)&=&0.27000, \nonumber\\
p_{X_e}(30)&=&0.28649,
\eear

and
\bear{ss0022}
p_{Y_e}(8)&=&0.63407, \nonumber\\
p_{Y_e}(9)&=&0.36593.
\eear
\fi
It is easy to verify that $G=2$.
\iffalse
From \req{genc5555}, we have
\bear{ss0033}
%\ex[X]&=&4.169659059, \nonumber\\
%\ex[Y]&=&8.339056783.
\ex[X]&=&4.16966, \nonumber\\
\ex[Y]&=&8.33906.
\eear
Thus, $G=\ex[Y]/\ex[X]=2$.
From \req{genc3333} and \req{genc4444}, we have
\bear{ss0044}
%p_{X}(2)&=&0.548893911, \nonumber\\
%p_{X}(3)&=&0.250457517, \nonumber\\
%p_{X}(7)&=&0.160829704, \nonumber\\
%p_{X}(30)&=&0.039818854,
p_{X}(2)&=&0.54889, \nonumber\\
p_{X}(3)&=&0.25046, \nonumber\\
p_{X}(7)&=&0.16083, \nonumber\\
p_{X}(30)&=&0.03982,
\eear
and
\bear{ss0055}
%p_{Y}(8)&=&0.660963921, \nonumber\\
%p_{Y}(9)&=&0.339067404.
p_{Y}(8)&=&0.66096, \nonumber\\
p_{Y}(9)&=&0.33907.
\eear
\fi
It was shown in \cite{shokrollahi2000design} that the maximum tolerable erasure probability  $\delta^*=0.49553$. %and it is wthin less than 1\% of the optimum 0.4985 (the upper bound is $1/G=0.5$).

%{\color{red}
	\iffalse
In this numerical experiment, consider the following bivariate distribution:
	\bear{impr1111}
	&&\pr(X_e=2, Y_e=8)=p_{11}, \nonumber\\
	&&\pr(X_e=2, Y_e=9)=p_{12}, \nonumber\\
	&&\pr(X_e=3, Y_e=8)=p_{21}, \nonumber\\
	&&\pr(X_e=3, Y_e=9)=p_{22}, \nonumber\\
	&&\pr(X_e=7, Y_e=8)=p_{31}, \nonumber\\
	&&\pr(X_e=7, Y_e=9)=p_{32}, \nonumber\\
	&&\pr(X_e=30, Y_e=8)=p_{41}, \nonumber\\
	&&\pr(X_e=30, Y_e=9)=p_{42},
	\eear
where
$0 \le p_{11} \le p_{X_e}(2)$, $0 \le p_{21} \le p_{X_e}(3)$, $0 \le p_{31} \le p_{X_e}(7)$ and $p_{X_e}(2)+p_{X_e}(3)+p_{X_e}(7)-p_{Y_e}(9) \le p_{11}+p_{21}+p_{31} \le p_{Y_e}(8)$.
\fi
For this example, \iffalse we verify that the maximum tolerable erasure probability $\delta^*$ is 0.49553. Now \fi we find the bivariate distribution $\pr(X_e=2, Y_e=8)=0.1534$, $\pr(X_e=3, Y_e=8)=0.1789$, $\pr(X_e=7, Y_e=8)=0.1035$ such that it has the same marginal distributions $p_{X_e}(x)$ and $p_{Y_e}(y)$. The maximum tolerable erasure probability $\delta^*$ for such a bivariate distribution is 0.49568, which is larger than 0.49553 for the LDPC code in \cite{shokrollahi2000design}.

Another example is the LDPC code in Example 3.63 of the book \cite{richardson2008modern}. The maximum tolerable erasure probability $\delta^*$ for that LDPC code is 0.47410. Using the general construction in \rsec{general}, we can extend
$\delta^*$ to 0.48077. Further detail can be found in Appendix \ref{sec:appdetail}.

%Other results and further details can be found in the full report in ArXiv with the same title.

\section{Conclusion}
\label{sec:con}

In this paper, we proposed two constructions of degree-degree correlated LDPC codes and derived the density evolution equations for such LDPC codes. For the block construction, our numerical results show that adding a negative degree-degree correlation can achieve a much higher maximum tolerable erasure probability than LDPC codes with independent degree distributions. Moreover, adding a negative degree-degree correlation could lead to better designs of LDPC codes with the UEP property. The general construction (with an arbitrary bivariate degree-degree distribution) provides a much larger ensemble of LDPC codes than block construction. It can lead to performance improvement over the existing LDPC codes with independent degree-degree distributions.

Our future work will extend our (density evolution) analysis for degree-degree correlated LDPC codes to more general multi-edge type LDPC codes. This will include convolutional LDPC codes (see, e.g., \cite{kudekar2011threshold,mitchell2015spatially}) with multiple node types.

\bibliographystyle{IEEEtran}
\bibliography{grantfreeCapacity}

\clearpage

\appendices
%\section*{Appendix A}

%\setcounter{section}{1}

%In this section, we provide a list of notations used in this paper.

\bsec{An example of two types of degrees by using the general construction}{apptwodegreesb}

%\bex{twodegreesb}{(Two types of degrees)}
Consider the following bivariate distribution:
\bear{twodb1111}
&&\pr(X_e=d_1, Y_e=Gd_1)=p_{1}, \nonumber\\
&&\pr(X_e=d_1, Y_e=Gd_2)=p_{2}, \nonumber\\
&&\pr(X_e=d_2, Y_e=Gd_1)=p_{2}, \nonumber\\
&&\pr(X_e=d_2, Y_e=Gd_2)=1-p_1-2p_2,
\eear
where
$0 \le p_2 \le 1/2$ and $0 \le p_1 \le 1-2p_2$.
\iffalse
Then,
\bear{twodb2222}
&&p_{X_e}(d_1)=p_{Y_e}(Gd_1)=p_1+p_2, \nonumber\\
&&p_{X_e}(d_2)=p_{Y_e}(Gd_2)=1-p_1-p_2.
\eear
\fi

From \req{genc5555}, it follows that
\bear{twodb4444}
\ex[X]&=&\frac{1}{\frac{p_1+p_2}{d_1}+\frac{1-p_1-p_2}{d_2}},
\nonumber\\
\ex[Y]&=&\frac{1}{\frac{p_1+p_2}{Gd_1}+\frac{1-p_1-p_2}{Gd_2}}.
\eear
Thus, the condition in \req{const1155} is satisfied.
\iffalse
From \req{genc3333} and \req{genc4444}, it follows that
\bear{twodb3333}
&&p_X(d_1)=p_Y(Gd_1)=\frac{\frac{p_1+p_2}{d_1}}{\frac{p_1+p_2}{d_1}+\frac{1-p_1-p_2}{d_2}}, \nonumber\\
&&p_X(d_2)=p_Y(Gd_2)=\frac{\frac{1-p_1-p_2}{d_2}}{\frac{p_1+p_2}{d_1}+\frac{1-p_1-p_2}{d_2}}.
\eear
\fi

From \req{twodb1111}\iffalse and \req{twodb2222}\fi, we know that
\bear{twodb5555}
\pr (Y_e=Gd_2| X_e=d_2)&=&\frac{1-p_1-2p_2}{1-p_1-p_2}, \nonumber\\
\pr (Y_e=Gd_2| X_e=d_1)&=&\frac{p_2}{p_1+p_2}, \nonumber\\
\pr (Y_e=Gd_1| X_e=d_2)&=&\frac{p_2}{1-p_1-p_2}, \nonumber\\
\pr (Y_e=Gd_1| X_e=d_1)&=&\frac{p_1}{p_1+p_2} .
\eear
It is easy to see that \req{twodb5555} recovers \req{twod1111} by letting $p_1=\frac{1-q}{4}$, $p_2=\frac{1+q}{4}$, $d_1=2d$ and $d_2=d$.
As such, the construction in this example is more general than that in \rex{twodegrees}.

Similarly,
from \req{twodb1111}\iffalse and \req{twodb2222} \fi, it follows that
\bear{twodb6666}
\pr (X_e=d_2| Y_e=Gd_2)&=&\frac{1-p_1-2p_2}{1-p_1-p_2}, \nonumber\\
\pr (X_e=d_2| Y_e=Gd_1)&=&\frac{p_2}{p_1+p_2}, \nonumber\\
\pr (X_e=d_1| Y_e=Gd_2)&=&\frac{p_2}{1-p_1-p_2} , \nonumber\\
\pr (X_e=d_1| Y_e=Gd_1)&=&\frac{p_1}{p_1+p_2} .
\eear
%\eex

\begin{figure}[ht]
	\centering
	\includegraphics[width=0.43\textwidth]{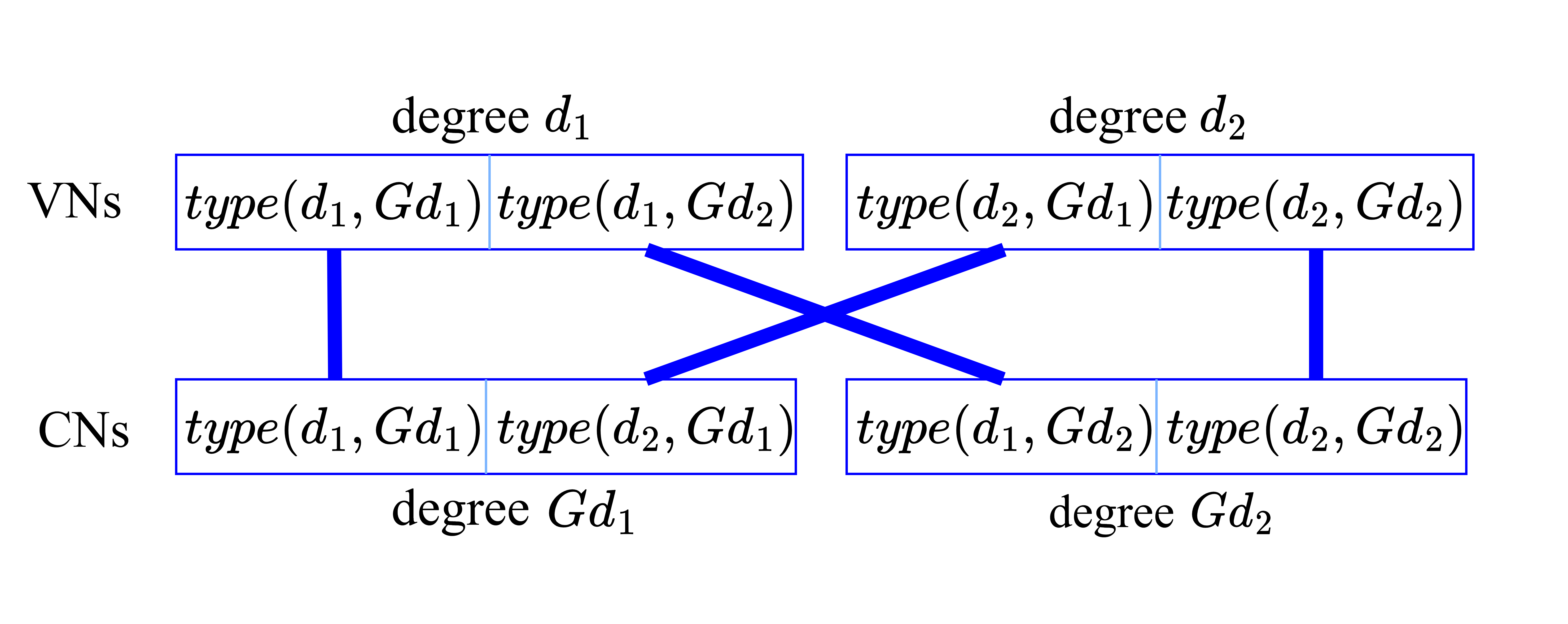}
	\caption{An illustration of the construction of the bipartite graph, where the bold lines represent edges connected among the four edge types.}
	\label{fig:stubgen}
\end{figure}

To construct a random LDPC code with the six input parameters $p_1$, $p_2$, $d_1$, $d_2$, $G$, and $n$ in this example, one can classify the $n \ex[X]$ edges (with $\ex[X]$ in \req{twodb4444}) into the four edge types $(x,y)$ (with degree $x$ in the variable end and degree $y$ in the check end): $(d_1, Gd_1)$, $(d_1,Gd_2)$, $(d_2, Gd_1)$ and $(d_2, Gd_2)$. For each edge type, connect the stubs in the variable nodes and the stubs in check nodes by using the configuration model. An illustration of the construction is shown in \rfig{stubgen}, where the bold lines represent edges connected among the four edge types.

\bsec{Proof of \rthe{stability}}{appproof}

Note from \req{andor2222c} that
\beq{stab2222}
\beta_y^{(i)} \le 1.
\eeq
Since $\alpha_x^{(0)}=\delta$, one can easily show by induction (from \req{andor2222c} and \req{andor1111c}) that
for all $i$
\beq{stab1111}
\alpha_x^{(i)} \le \delta.
\eeq
Let $\alpha_{\max}^{(i)}=\max_x \alpha_x^{(i)} $ and $\beta_{\max}^{(i)}=\max_y \beta_y^{(i)} $.
Using the inequality that $(1-z)^k \ge 1-kz$ for all $z \ge 0$, we have from \req{andor2222c} that
\bear{stab3333}
\beta_y^{(i)}&\le& (y-1)\sum_x \alpha_x^{(i-1)} \pr(X_e=x|Y_e=y) \nonumber\\
&\le& (y-1)\alpha_{\max}^{(i-1)}\sum_x  \pr(X_e=x|Y_e=y) \nonumber\\
&\le& (d_{c,\max}-1)\alpha_{\max}^{(i-1)}.
\eear
It follows from \req{stab1111} that
\beq{stab4444}
\beta_{\max}^{(i)} \le (d_{c,\max}-1)\alpha_{\max}^{(i-1)} \le (d_{c,\max}-1)\delta.
\eeq
On the other hand, we have from \req{andor1111c} and \req{stab4444} that
\bear{stab5555}
\alpha_x^{(i)}&\le& \delta (\beta_{\max}^{(i)}\sum_y  \pr(Y_e=y |X_e=x))^{x-1}  \nonumber\\
&\le&\delta (\beta_{\max}^{(i)})^{x-2} \beta_{\max}^{(i)} \nonumber\\
&\le& \delta((d_{c,\max}-1)\delta)^{x-2}  (d_{c,\max}-1)\alpha_{\max}^{(i-1)}  \nonumber\\
&\le &((d_{c,\max}-1)\delta)^{x-1} \alpha_{\max}^{(i-1)}
\eear
Since we assume that $(d_{c,\max}-1)\delta <1$, it follows from \req{stab5555}
that
\beq{stab6666}
\alpha_x^{(i)} \le ((d_{c,\max}-1)\delta)^{d_{v,\min}-1} \alpha_{\max}^{(i-1)} .
\eeq
Thus,
\beq{stab7777}
\alpha_{\max}^{(i)} \le ((d_{c,\max}-1)\delta)^{d_{v,\min}-1} \alpha_{\max}^{(i-1)} .
\eeq
Since $d_{v,\min} \ge 2$ and  $(d_{c,\max}-1)\delta <1$, we have from \req{stab7777} and $\alpha_x^{(0)}=\delta$ that
\beq{stab8888}
\alpha_{\max}^{(i)} \le  (((d_{c,\max}-1)\delta)^{d_{v,\min}-1})^{i}\delta
.
\eeq
Thus, $\alpha_{\max}^{(i)}$ converges to 0 when $i \to \infty$.

\bsec{Simulation results for the block construction in \rsec{blocknum}}{appsimulation}

\begin{figure}[ht]
	\centering
	\includegraphics[width=0.43\textwidth]{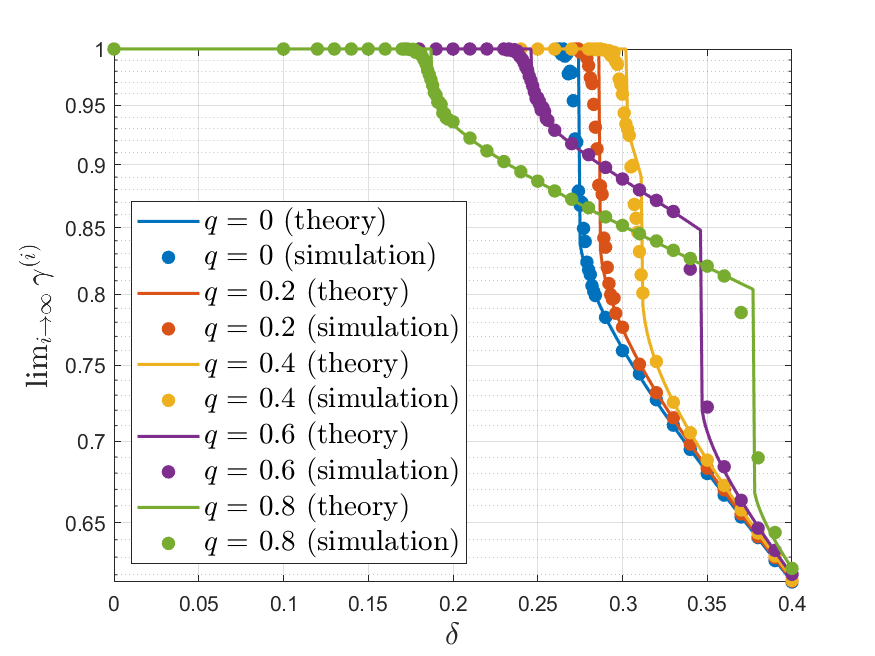}
	\caption{The probability that a randomly selected variable node is successfully decoded, i.e., $\lim_{i \to \infty} \gamma^{(i)}$, as a function of the erasure probability $\delta$ for $q = 0,0.2,0.4,0.6,0.8.$}
	\label{fig:ex111}
\end{figure}

In order to verify the effectiveness of the asymptotic results derived from the density evolution equations,
%in \req{twod2222} and \req{twod2244}
we conduct extensive simulations. For our simulations, we generate degree-degree correlated LDPC codes with $n=18,000$ variable nodes and $n-k=6,000$ check nodes by using the construction in \rex{twodegrees}. Then we simulate the LDPC codes over a BEC with an erasure probability $\delta$.
After that, we perform iterative decoding (peeling decoder) for the LDPC codes (until there are no check nodes with degree $1$). In \rfig{ex111}, we plot the probability that a randomly selected variable node is successfully decoded, i.e., $\lim_{i \to \infty} \gamma^{(i)}$, as a function of the erasure probability $\delta$ from 0 to 0.4 with respect to various parameters $q=0, 0.2, 0.4, 0.6, 0.8$, respectively.
Each data point is the average of 100 experiments.
The five solid curves represent the theoretical results of $\lim_{i \to \infty} \gamma^{(i)}$, and the five dotted curves represent the corresponding simulation results.
As shown in \rfig{ex111}, the simulation results and the asymptotic results match very well.
Also, we observe the largest maximum tolerable erasure probability occurs when $q=0.4$ (the yellow curve). This result is consistent with that of \rfig{mtep}.

\begin{figure}[ht]
	\centering
	\includegraphics[width=0.43\textwidth]{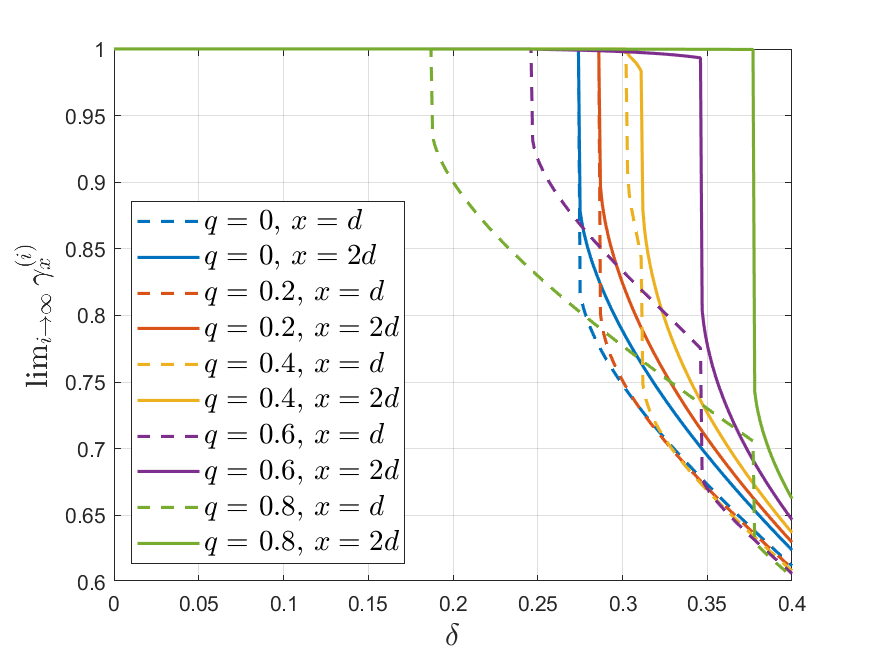}
	\caption{The probability that a randomly selected variable node with degree $x$ (for $x=d, 2d$) is successfully decoded, i.e.,  $\lim_{i \to \infty} \gamma^{(i)}_x$,  as a function of the erasure probability $\delta$ for $q = 0, 0.2, 0.4, 0.6, 0.8$.}
	\label{fig:uu111}
\end{figure}

Finally, we look into the effect of the degree on the decoding probabilities of variable nodes.
%{\color{red} from the density evolution equations in \req{twod2233}}.
There are two types of degrees of the variable nodes in \rex{twodegrees}: $d$ and $2d$. In \rfig{uu111}, we plot the probability that a randomly selected variable node with degree $x$ is successfully decoded, i.e.,  $\lim_{i \to \infty} \gamma^{(i)}_x$,  as a function of the erasure probability $\delta$ for $q = 0, 0.2, 0.4, 0.6, 0.8$.  The solid curve (resp. dashed curve) represents the decoding probability for a variable node with degree $d$ (resp. $2d$).
It is well-known that the decoding probability for variable nodes with a large degree is higher than that with a small degree. Such a property is known as  the unequal error protection (UEP) property.
One significant finding from our experiments is that the degree-degree correlation  plays a critical  role in UEP.
Recall that the  degree-degree correlation in \rex{twodegrees} is $-q$.
As shown in \rfig{uu111}, increasing $q$ increases the gap between the curve for $\lim_{i \to \infty} \gamma^{(i)}_d$ and the curve for $\lim_{i \to \infty} \gamma^{(i)}_{2d}$. In particular, for $q=0.8$ (the two green curves), the gap is the largest among the five choices of $q$. Even when the erasure probability exceeds the upper bound $1/G=1/3$,
variable nodes with degree $2d$ can still have a very high decoding probability. This is at the cost of sacrificing the decoding probability of variable nodes with degree $d$.
On the other hand, the two curves for $q=0.4$ are very close to each other. Even though they have the largest maximum tolerable erasure probability $\delta^*$, they are not suitable for LDPC codes with the UEP property.

\bsec{Further detail for performance improvement}{appdetail}

In this appendix, we provide the details of the performance improvement in \rsec{generalnum}.

%show that the LDPC codes from the general construction in \rsec{general} can lead to further performance improvement.

%\iffalse
\subsubsection{The LDPC code by Shokrollahi and Storn \cite{shokrollahi2000design}}

In \cite{shokrollahi2000design}, Shokrollahi and Storn proposed
a construction of the LDPC code with the following (independent) bivariate distribution:
\beq{ss0000app}
\pr (X_e=x, Y_e=y)=p_{X_e}(x)p_{Y_e}(y),
\eeq
where
\bear{ss0011}
p_{X_e}(2)&=&0.26328, \nonumber\\
p_{X_e}(3)&=&0.18020, \nonumber\\
p_{X_e}(7)&=&0.27000, \nonumber\\
p_{X_e}(30)&=&0.28649,
\eear
and
\bear{ss0022}
p_{Y_e}(8)&=&0.63407
, \nonumber\\
p_{Y_e}(9)&=&0.36593.
\eear
From \req{genc5555}, we have
\bear{ss0033}
\ex[X]&=&4.16966, \nonumber\\
\ex[Y]&=&8.33906.
\eear
\iffalse
\bear{ss0033}
\ex[X]&=&4.169659059, \nonumber\\
\ex[Y]&=&8.339056783.
\eear
\fi
Thus, $G=\ex[Y]/\ex[X]=2$.
From \req{genc3333} and \req{genc4444}, we have
\bear{ss0044}
p_{X}(2)&=&0.54889, \nonumber\\
p_{X}(3)&=&0.25046, \nonumber\\
p_{X}(7)&=&0.16083, \nonumber\\
p_{X}(30)&=&0.03982,
\eear
\iffalse
\bear{ss0044}
p_{X}(2)&=&0.548893911, \nonumber\\
p_{X}(3)&=&0.250457517, \nonumber\\
p_{X}(7)&=&0.160829704, \nonumber\\
p_{X}(30)&=&0.039818854,
\eear
\fi
and
\bear{ss0055}
p_{Y}(8)&=&0.66096, \nonumber\\
p_{Y}(9)&=&0.33907.
\eear
\iffalse
\bear{ss0055}
p_{Y}(8)&=&0.660963921, \nonumber\\
p_{Y}(9)&=&0.339067404.
\eear
\fi
It was shown in \cite{shokrollahi2000design} that $\delta^*=0.4955$ and it is wthin less than 1\% of the optimum 0.4985 (the upper bound is $1/G=0.5$).

\iffalse
Things to do: (i) Verify that $\delta^*=0.4955$ in this example. (ii) Find a bivariate distribution with the same marginal distributions $p_{X_e}(x)$ in \req{ss0011}  and $p_{Y_e}(y)$ in \req{ss0022} such that $\delta^*$ is larger than 0.4955.
For the second part, it might be better to write a differential evolution algorithm (for the search). Grid search is very difficult as there are too many parameters. Use the key words ``differential evolution'' for a Google search to find the webpage (and the tutorial videos) for the differential evolution algorithm.
\fi

In this numerical experiment, we conduct a search for the following set of bivariate distributions:
	\bear{impr1111a}
	&&\pr(X_e=2, Y_e=8)=p_{11}, \nonumber\\
	&&\pr(X_e=2, Y_e=9)=p_{12}, \nonumber\\
	&&\pr(X_e=3, Y_e=8)=p_{21}, \nonumber\\
	&&\pr(X_e=3, Y_e=9)=p_{22}, \nonumber\\
	&&\pr(X_e=7, Y_e=8)=p_{31}, \nonumber\\
	&&\pr(X_e=7, Y_e=9)=p_{32}, \nonumber\\
	&&\pr(X_e=30, Y_e=8)=p_{41}, \nonumber\\
	&&\pr(X_e=30, Y_e=9)=p_{42},
	\eear
where
$0 \le p_{11} \le p_{X_e}(2)$, $0 \le p_{21} \le p_{X_e}(3)$, $0 \le p_{31} \le p_{X_e}(7)$ and $p_{X_e}(2)+p_{X_e}(3)+p_{X_e}(7)-p_{Y_e}(9) \le p_{11}+p_{21}+p_{31} \le p_{Y_e}(8)$.
For this example, we verify that the maximum tolerable erasure probability is $\delta^*=0.49553$ with $p_{11}=0.1669$, $p_{21}=0.1143$, $p_{31}=0.1712$. Now we find the bivariate distribution $p_{11}=0.1534$, $p_{21}=0.1789$, $p_{31}=0.1035$ such that it has the same
marginal distributions $p_{X_e}(x)$ in \req{ss0011} and $p_{Y_e}(y)$ in \req{ss0022}. The maximum tolerable erasure probability $\delta^*$ for such a bivariate distribution is 0.49568, which is larger than 0.49553 for the LDPC code
in \cite{shokrollahi2000design}.

\subsubsection{The LDPC code in Example 3.63 of the book \cite{richardson2008modern}}

In Example 3.63 of the book \cite{richardson2008modern},  the LDPC code with the following (independent) bivariate distribution is considered:
\beq{modern0000}
\pr (X_e=x, Y_e=y)=p_{X_e}(x)p_{Y_e}(y),
\eeq
where
\bear{modern0011}
p_{X_e}(2)&=&0.10626, \nonumber\\
p_{X_e}(3)&=&0.48666, \nonumber\\
p_{X_e}(11)&=&0.01039, \nonumber\\
p_{X_e}(20)&=&0.39669,
\eear
\iffalse
\bear{modern0011}
p_{X_e}(2)&=&0.106257, \nonumber\\
p_{X_e}(3)&=&0.486659, \nonumber\\
p_{X_e}(11)&=&0.010390, \nonumber\\
p_{X_e}(20)&=&0.396694,
\eear
\fi
and
\bear{modern0022}
p_{Y_e}(8)&=&0.5
, \nonumber\\
p_{Y_e}(9)&=&0.5.
\eear
It is easy to verify that $G=2$. For such an LDPC code, $\delta^*=0.4741$

	In this numerical experiment, we conduct a search for the following set of bivariate distributions:
	\bear{impr1111b}
	&&\pr(X_e=2, Y_e=8)=p_{11}, \nonumber\\
	&&\pr(X_e=2, Y_e=9)=p_{12}, \nonumber\\
	&&\pr(X_e=3, Y_e=8)=p_{21}, \nonumber\\
	&&\pr(X_e=3, Y_e=9)=p_{22}, \nonumber\\
	&&\pr(X_e=11, Y_e=8)=p_{31}, \nonumber\\
	&&\pr(X_e=11, Y_e=9)=p_{32}, \nonumber\\
	&&\pr(X_e=20, Y_e=8)=p_{41}, \nonumber\\
	&&\pr(X_e=20, Y_e=9)=p_{42},
	\eear
	where
	$0 \le p_{11} \le p_{X_e}(2)$, $0 \le p_{21} \le p_{X_e}(3)$, $0 \le p_{31} \le p_{X_e}(11)$ and $p_{X_e}(2)+p_{X_e}(3)+p_{X_e}(11)-p_{Y_e}(9) \le p_{11}+p_{21}+p_{31} \le p_{Y_e}(8)$.
	%In this example, we verify that the maximum tolerable erasure probability is $\delta^*=0.4741$ with parameters $p_{11}=0.0531$, $p_{21}=0.2433$, $p_{31}=0.0052$. Now we take a bivariate distribution $p_{11}=0$, $p_{21}=0.274$, $p_{31}=0.007$ with the same marginal distributions $p_{X_e}(x)$ in \req{ss0011} and $p_{Y_e}(y)$ in \req{ss0022} to extend the maximum tolerable erasure probability $\delta^*$ from 0.4741 to 0.48077. 	
	For this example, we verify that the maximum tolerable erasure probability is $\delta^*=0.47410$ with $p_{11}=0.0531$, $p_{21}=0.2433$, $p_{31}=0.0052$. Now we find the bivariate distribution $p_{11}=0$, $p_{21}=0.274$, $p_{31}=0.007$ such that it has the same marginal distributions $p_{X_e}(x)$ in \req{ss0011} and $p_{Y_e}(y)$ in \req{ss0022}. The maximum tolerable erasure probability $\delta^*$ for such a bivariate distribution is 0.48077, which is larger than 0.47410 for the LDPC code
	in \cite{richardson2008modern}.

\subsubsection{The LDPC code by Bazzi et al. \cite{bazzi2004exact}}

%\bex{bazz}{(Bazzi et al.  (2004))}
In Example 9 of \cite{bazzi2004exact}, Bazzi et al. considered the rate one-half
LDPC code with the following (independent) bivariate distribution:
\beq{bazz0000}
\pr (X_e=x, Y_e=y)=p_{X_e}(x)p_{Y_e}(y),
\eeq
where
\bear{bazz0011}
p_{X_e}(3)&=&a, \nonumber\\
p_{X_e}(4)&=&1-a,
\eear
and
\bear{bazz0022}
p_{Y_e}(7)&=&\frac{7a}{3}
, \nonumber\\
p_{Y_e}(8)&=&1-\frac{7a}{3},
\eear
where $a$ is chosen to be 0.1115.
From \req{genc5555}, we have
\bear{bazz0033}
\ex[X]&=&\frac{12}{a+3}, \nonumber\\
\ex[Y]&=&\frac{24}{a+3}.
\eear
Thus, $G=\ex[Y]/\ex[X]=2$.
From \req{genc3333} and \req{genc4444}, we have
\bear{bazz0044}
p_{X}(3)&=&\frac{4a}{a+3}
, \nonumber\\
p_{X}(4)&=&1-\frac{4a}{a+3}
,
\eear
and
\bear{bazz0055}
p_{Y}(7)&=&\frac{8a}{a+3}
, \nonumber\\
p_{Y}(8)&=&1-\frac{8a}{a+3}.
\eear
%\eex

In this numerical experiment, we conduct a search for the following set of bivariate distributions:
\bear{impr1111c}
&&\pr(X_e=3, Y_e=7)=p_{11}, \nonumber\\
&&\pr(X_e=3, Y_e=8)=p_{12}, \nonumber\\
&&\pr(X_e=4, Y_e=7)=p_{21}, \nonumber\\
&&\pr(X_e=4, Y_e=8)=p_{22},
\eear
where
$0 \le p_{11} \le p_{X_e}(3)$.
For this example with $a=0.1115$, we verify that the maximum tolerable erasure probability is $\delta^*=0.3916$ with  $p_{11}=0.029$. Now we find the bivariate distribution $p_{11}=0$ such that it has the same marginal distributions $p_{X_e}(x)$ in \req{ss0011} and $p_{Y_e}(y)$ in \req{ss0022}. The maximum tolerable erasure probability $\delta^*$ for such a bivariate distribution is 0.3924, which is larger than 0.3916 for the LDPC code in \cite{bazzi2004exact}.

\end{document}